\documentclass[pra,aps,amsfonts,amsmath,superscriptaddress,twocolumn,showpacs,floatfix]{revtex4}
\usepackage{amsfonts}
\usepackage{amsmath}
\usepackage{bm}
\usepackage{epsfig}
\usepackage{ulem}
\usepackage{amssymb}
\usepackage{graphicx}%
\providecommand{\U}[1]{\protect\rule{.1in}{.1in}}

\begin{document}
\title{\textbf{Nuclear structure in strong magnetic fields: nuclei in the crust
of a
magnetar}}
\author{D. Pe\~na Arteaga}
\affiliation{Institut de Physique Nucl\'eaire, Universit\'e Paris-Sud,
IN2P3-CNRS, F-91406
Orsay Cedex, France}
\author{M. Grasso}
\affiliation{Institut de Physique Nucl\'eaire, Universit\'e Paris-Sud,
IN2P3-CNRS, F-91406
Orsay Cedex, France}
\author{E. Khan}
\affiliation{Institut de Physique Nucl\'eaire, Universit\'e Paris-Sud,
IN2P3-CNRS, F-91406
Orsay Cedex, France}
\author{P. Ring}
\affiliation{Physikdepartment, Technische Universit\"{a}t M\"{u}nchen, D-85748,
Garching, Germany}

\begin{abstract}
Covariant density functional theory is used to study the effect of strong
magnetic fields, up to the limit predicted for neutron stars (for magnetars $B
\approx10^{18}$G), on nuclear structure. All new terms in the equation of
motion resulting from time reversal symmetry breaking by the magnetic field
and the induced currents, as well as axial deformation, are taken into account
in a self-consistent fashion. For nuclei in the iron region of the nuclear
chart it is found that fields in the order of magnitude of $10^{17}$G
significantly affect bulk properties like masses and radii.

\end{abstract}
\maketitle

\section{Introduction}

Several studies (e.g. \cite{Shapiro1983,Haensel2006}) have determined the
presence of intense magnetic fields, up to $10^{16}$ G, on the surface of
neutron stars.  Theoretical models suggest that these magnetic fields might
reach up to $B \approx10^{18}$ G, and even larger values if one considers the
limit imposed by the virial theorem ($B \approx2 \cdot10^{18}$ G
\cite{Broderick2000}). The influence of the magnetic fields on the equation of
state (EOS) has been thoroughly studied and reported in last years (e.g.
\cite{Broderick2000,Guang-Jun2003,Wei2006,Rabhi2008}). It should be noted
lower than $10^{18}$~G magnetic fields can influence the low density parts of 
a neutron star, such as its surface layer. 


The outer crust is fundamentally composed of well separated nuclides, and its
structure determined by the energies of isolated nuclei, the kinetic energy of
electrons and the lattice energy. Thus its composition depends very much on the
binding energy of stable and unstable nuclei in the outer crust below the
neutron drip density and of neutron-rich nuclear systems above the neutron
drip. At the lowest densities it is thought that the most abundant component is
$^{56}$Fe because of its high binding energy, a fact actually observed in
emission spectra (e.g. \cite{Cackett2008}).

The magnetic field strength required to directly influence the EOS can be
estimated by considering its effects on charged particles. Charge-neutral,
beta-equilibrated neutron star matter contains both negatively charged leptons
(electrons and muons) and protons. Magnetic fields quantize the orbital motion
(Landau quantization) of these charged particles. When the Fermi energy of the
proton becomes significantly affected by the magnetic field, the composition of
matter in beta equilibrium is modified. This is reflected in a change of the
pressure of matter. It has been found in Ref.~\cite{Broderick2000} that this
occurs for fields of approximately $10^{18}$ G, and that in general leads to a
stiffening of the EOS.

However, there are very few studies \cite{Kondratyev2000,Kondratyev2001} of the
changes that these very intense magnetic fields may eventually bring to the
composition of the crust. The structure and composition of the crust is
important in the thermal and rotational evolution of neutron stars, in
particular in the theory behind glitches \cite{Shapiro1983}. Some other studies
\cite{Duncan1997,Thompson2001} point to a magnetically driven crust activity as
the source of soft gamma repeaters (SGR).

So far, the impact of intense magnetic fields in nuclei found in the outer
crust has been studied on a qualitative way using a simple non self-consistent
method by Kondratyev and collaborators \cite{Kondratyev2000,Kondratyev2001}. It
was found that fields as low as $10^{16}$ G may modify the nuclear shell
structure, well within the range of theoretically possible magnetic strengths.
There are, however, still some questions that need to be addressed: i) What is
the minimum field that is able to significantly alter the nuclear structure?
ii) Is this field low enough to be found in a significant proportion of neutron
stars or magnetars?  iii) Is this effect big enough to influence
astrophysically relevant situations and processes, e.g. neutron star outer
crust composition or final element abundances in nucleosynthesis scenarios?

The objective of the present work is to try to find answers to these
questions, in a quantitative way if possible, using a fully microscopical
description of the nuclear system within covariant density functional theory.
The formalism used will be introduced in Section II. A general discussion of
the effects of the external magnetic field on nuclei will be given on Section
III, and particularized to an example nucleus. A discussion of the possible
influence of the magnetic fields on neutron star outer crust nuclei can be
found on Section IV. Finally, Section V is devoted to the conclusions.

\section{Formalism}

Covariant density functional theory starts from an effective Lagrangian, that
includes the nucleon and as many meson fields as needed to reproduce basic
nuclear properties like saturation (a detailed discussion can be found in
Refs.~\cite{Gambhir1990,Vretenar2005} ):
\begin{equation}
\mathcal{L}=\mathcal{L}_{N}+\mathcal{L}_{m}+\mathcal{L}_{int}+\mathcal{L}%
_{BO}+\mathcal{L}_{BM}.\label{e:lagdens}%
\end{equation}
$\mathcal{L}_{N}$ refers to the Lagrangian of the free nucleon
\begin{equation}
\mathcal{L}_{N}=\bar{\psi}(i{\gamma}^{\mu}{\partial}_{\mu}-m){\psi},
\end{equation}
where $m$ is the bare nucleon mass and ${\psi}$ denotes the Dirac spinor.
$\mathcal{L}_{m}$ is the Lagrangian of the free meson fields and the
electromagnetic field generated by the protons
\begin{align}
\mathcal{L}_{m} &  =\frac{1}{2}\partial_{\mu}\sigma\partial^{\mu}\sigma
-\frac{1}{2}m_{\sigma}^{2}{\sigma}^{2}-\frac{1}{4}\Omega_{\mu\nu}{\Omega}%
^{\mu\nu}+\frac{1}{2}m_{\omega}^{2}{\omega}_{\mu}{\omega}^{\mu}\nonumber\\
&  -\frac{1}{4}\vec{R}_{\mu\nu}\vec{R}^{\mu\nu}+\frac{1}{2}m_{\rho}^{2}%
\vec{\rho}_{\mu}\vec{\rho}^{\mu}-\frac{1}{4}F_{\mu\nu}F^{\mu\nu}+U(\sigma),
\end{align}
where $m_{\sigma}$, $m_{\omega}$, $m_{\rho}$ are the meson masses, and
$U(\sigma)=(g_{2}/3)\sigma^{3}+(g_{3}/4)\sigma^{4}$ is the standard form for the
non-linear coupling of the $\sigma$ meson field. The interaction Lagrangian
$\mathcal{L}_{int}$ is given by minimal coupling terms
\begin{align}
\mathcal{L}_{int}= &  -g_{\sigma}\bar{\psi}\sigma\psi-g_{\omega}\bar{\psi
}\gamma^{\mu}\omega_{\mu}\psi\nonumber\\
&  -g_{\rho}\bar{\psi}\gamma^{\mu}\vec{\tau}\vec{\rho}_{\mu}\psi-e\bar{\psi
}\gamma^{\mu}A_{\mu}\psi,
\end{align}
where $g_{\sigma}$, $g_{\omega}$, $g_{\rho}$ and $e$ are the respective
coupling constants for the ${\sigma}$, ${\omega}$, $\vec{\rho}$ and photon
fields and, of course, $e$ vanishes for neutrons. In the previous and
subsequent formulae, bold symbols denote vectors in ordinary space, and arrows
vectors in isospin space. These three terms, $\mathcal{L}_{N}$, $\mathcal{L}%
_{m}$ and $\mathcal{L}_{int}$ compose the standard RMF Lagrangian. Throughout
this work the parameter set NL3~\cite{Lalazissis1997} is used for the
masses and coupling
constants of the model. This parameter set has been thoroughly used and has led to 
very successful description of many nuclear properties. In addition, there are 
two new terms corresponding to
the interaction of the nuclear system with an external magnetic field: i) the
coupling of the proton orbital motion with the external magnetic field,
\begin{equation}
\mathcal{L}_{BO}=-e\bar{\psi}\gamma^{\mu}A_{\mu}^{(e)}\psi,\label{BO}%
\end{equation}
and ii) the coupling of both proton and neutron intrinsic dipole magnetic
moments with the external magnetic field \cite{Bjorken1964}
\begin{equation}
\mathcal{L}_{BM}=-\bar{\psi}\chi_{\tau_{3}}^{(e)}\psi,\label{BM}%
\end{equation}
where
\begin{equation}
\chi_{\tau_{3}}^{(e)}=\kappa_{\tau_{3}}\mu_{N}\frac{1}{2}\sigma_{\mu\nu
}F^{(e)\mu\nu}.
\end{equation}
Here $\sigma_{\mu\nu}=\frac{i}{2}\left[
\gamma_{\mu},\gamma_{\nu}\right]  $ and $\mu_{N}=e\hslash/2m$ is the
nuclear magneton and $\kappa_{n}=g_{n}/2$, $\kappa_{p}=g_{p}/2-1$
with $g_{n}=-3.8263$ and $g_{p}=5.5856$ are the intrinsic magnetic
moments of protons and neutrons. Interactions with the external
magnetic field are marked the superscript (e). This field is
considered to be externally generated, and therefore there is no associated
field equation and thus no other bosonic terms in the Lagrangian.
Both terms $\mathcal{L}_{BO}$ (\ref{BO}) and $\mathcal{L}_{BM}$
(\ref{BM}) have to be taken into account since at the magnetic field
strengths of interest ($B\approx10^{17}$G) they are of the same order
of magnitude.

The Hamiltonian density can be derived from the Lagrangian density of
Eq.~\eqref{e:lagdens} as the (0,0) component of the energy-momentum tensor,
leading the to the energy functional $E_{\boldsymbol{B}}[\hat{\rho},\phi]$
(see in Ref.~\cite{Vretenar2005} for details).
\begin{align}
E_{\boldsymbol{B}}[\hat{\rho},\phi]  &  = \mathrm{Tr}\left[  \left(
\boldsymbol{\alpha} \left(  -i\boldsymbol{\nabla} - e\boldsymbol{A}^{(e)}
\right)  + \beta( m + \chi^{(e)}_{\tau_{3}} ) \right)  \hat{\rho}\right]
\nonumber\\
&  +\sum_{m}\mathrm{Tr}\left[  \left(  \beta\,\boldsymbol{\Gamma}_{m}\phi
_{m}\right)  \hat{\rho}\right] \nonumber\\
&  \pm\sum_{m}{\int}d^{3}r \left[  \frac{1}{2}({\partial}_{\mu}{\phi}_{m}%
)^{2}+\frac{1}{2}m^{2}_{m}\phi^{2}_{m}\right]  , \label{e:efunctional}%
\end{align}
where the upper sign holds for scalar and the lower sign for vector mesons.
\begin{equation}
\hat{\rho}(\boldsymbol{r})=\sum_{i} \vert\psi_{i}(\boldsymbol{r}%
)\rangle\langle\psi_{i}(\boldsymbol{r})\vert
\end{equation}
is the relativistic single-particle density matrix and the traces run over the
Dirac indices and over the integral in $r$-pace and, according to the no-sea
approximation, the index $i$ runs over all the occupied levels in the Fermi
sea. The index $m=\sigma,\omega,\rho,e$ runs over the various meson and
electromagnetic fields and the vertices $\Gamma_{m}$ read
\begin{align}
\Gamma_{\sigma} = g_{\sigma},~~~~\,  &  \qquad\Gamma^{\mu}_{\omega} =
g_{\omega}\gamma^{\mu},\nonumber\\
\vec{\Gamma}^{\mu}_{\rho} = g_{\rho} \vec{\tau}\gamma^{\mu},  &  \qquad
\Gamma^{\mu}_{e} = e\gamma^{\mu}, \label{rmf:Lvertex}%
\end{align}
and $\frac{1}{2}m^{2}_{m}\phi^{2}_{m}$ has to be replaced by $\frac{1}{2}%
m^{2}_{\sigma}\sigma^{2}+U(\sigma)$ in the case of the $\sigma$-meson. It is
customary at this point to introduce an additional term into the energy
functional to account for pairing correlations~\cite{Kucharek1991}, at least in
its
simplest BCS approximation. In the present study, however, pairing effects
shall be completely neglected. It is a well known fact \cite{Kittel1987} that
static magnetic fields lead to a reduction in pair correlations in
superconductors, and to the appearance of a critical field where all such
correlations vanish.

The functional (\ref{e:efunctional}) follows the spirit of
magnetic-field-and-density functional theory (BDFT) \cite{Grayce1994}, in which
the
vector potential is introduced as an explicit dependence in the energy
functional. Considering that astrophysical magnetic fields can be taken as
constant on a nuclear scale (i.e. their functional form is fixed), it would be
of little advantage to use the more general current-and-density functional
theory (CDFT) \cite{Vignale1990235}, which generalizes density functional theory
(DFT)
with the inclusion of an external vector potential in a universal fashion. And
since there is no practical value in considering the external magnetic field
$\boldsymbol{B}$ as an independent variable, it shall be regarded in the
density functional as a parametric variable. Minimization with respect to the
density $\hat{\rho}$ in the Hartree approximation \cite{Vretenar2005} and
considering only static configurations leads to the stationary Dirac equation
for the nucleons and to the Klein-Gordon equations for the mesons
\begin{align}
\hat{h}_{D}\psi_{i}  &  =\epsilon_{i}\psi_{i},\label{direq}\\
\left[  -\Delta+m_{m}^{2}\right]  \phi_{m}  &  =\mp\sum_{i}\bar{\psi}%
_{i}\Gamma_{m}\psi_{i}, \label{meseq}%
\end{align}
where $m_{m}^{2}$ has to be replaced by $m_{\sigma}^{2}+g_{2}\sigma
+g_{3}\sigma^{2}$ in the case of the $\sigma$-meson and where the Dirac
Hamiltonian has the form
\begin{align}
\hat{h}_{D}  &  =\frac{\delta E_{B}[\hat{\rho},\phi]}{\delta\hat{\rho}%
}\nonumber\\
&  =\boldsymbol{\alpha}\left(  -i\boldsymbol{\nabla}-\mathbf{V}\right) +V_{0}%
+\beta(m+S)+ \beta\chi_{\tau_{3}}^{(e)},
\end{align}
with the scalar and vector potentials $S$ and $V_{\mu}$ defined as
\begin{align}
S  &  =-g_{\sigma}\sigma,\\
V_{\mu}  &  =g_{\omega}\omega_{\mu}+g_{\rho}\tau_{3}\rho_{\mu,3}+eA_{\mu
}+eA_{\mu}^{(e)}, \label{e:14}%
\end{align}
At this point it may be useful to fix the functional form of the
magnetic field. Of course, it is not constant throughout the neutron
star. However, the scale of changes is much larger than the
microscopic nuclear scale \cite{Broderick2000}. Thus, the magnetic field
$\boldsymbol{B}$ within each individual nucleus might be considered
constant. If one chooses the intrinsic $z$-axis in the direction of
this constant external magnetic field $\boldsymbol{B}=(0,0,B)$ and
cylindrical coordinates $(z,r,\varphi)$ the contribution of this
external field to the vector potential $\mathbf{V}$ can be written,
in the symmetric gauge, as
\begin{equation}
\boldsymbol{A}^{(e)}=-\frac{rB}{2}\boldsymbol{e}_{\varphi}, \label{A-ext}%
\end{equation}
where $\boldsymbol{e}_{\varphi}$ is the unit vector associated with the
azimuth angle $\varphi$ and $r$ is the distance from the symmetry axis. As
discussed in Eq.~(\ref{sigmaB}) of the Appendix B the contribution of the
intrinsic magnetic moments is given by%
\begin{equation}
\chi_{\tau_{3}}^{(e)}=-\kappa_{\tau_{3}}\mu_{N}\Sigma_{3}B
\end{equation}

It is clear that the presence of the magnetic field breaks spherical
symmetry for the Dirac and Klein-Gordon equations. Only axial
symmetry is preserved for fields of the form (\ref{A-ext}). As
discussed in Ref.~\cite{Gambhir1990} the spinor solutions in
Eq.~(\ref{direq}) can be written, in axial symmetry, as
\begin{equation}
|\psi_{i}(\boldsymbol{r})\rangle=\frac{1}{\sqrt{2\pi}}\left(
\begin{array}
[c]{c}%
f_{i}^{+}(r,z)e^{i(\Omega_{i}-1/2)\varphi}\\
f_{i}^{-}(r,z)e^{i(\Omega_{i}+1/2)\varphi}\\
ig_{i}^{+}(r,z)e^{i(\Omega_{i}-1/2)\varphi}\\
ig_{i}^{-}(r,z)e^{i(\Omega_{i}+1/2)\varphi}\\
\end{array}
\right)  \chi_{t_{i}(t)}. \label{def:wavef}%
\end{equation}
They are characterized by the angular momentum projection $\Omega$, the parity
$\pi$ and the isospin projection $t$. For even-even nuclei and in the absence
of an external magnetic field, according to Kramers rule, for each solution
$\psi_{i}$ with positive $\Omega_{i}$ there exists a time-reversed one with
the same energy, denoted by a bar, $\bar{\imath}:=\{\epsilon_{i},\,-\Omega
_{i},\,\pi_{i}\}$. However, time-reversal symmetry is broken by the magnetic
field, so the two-fold degeneracy is not present and one needs to consider
both solutions separately. This breaking of time-reversal symmetry in the
intrinsic frame leads to the appearance of time-odd mean fields and
non-vanishing currents which induce space-like components of the vector mesons
$\omega$ and $\rho$, usually referred as nuclear
magnetism~\cite{Hofmann1988,Koepf1990,Afanasjev2010a}. It is a great advantage
in relativistic
nuclear density functionals that these time-odd mean fields are determined by
the same coupling constants $g_{\omega}$ and $g_{\rho}$ as the well determined
time-even fields.

In non-relativistic nuclear density functionals such as Skyrme
~\cite{Engel1975,Dobaczewski1995} or Gogny~\cite{Egido1993} there are, in
principle, also relations connecting time-even and time-odd parts through
Galilean and gauge invariance~\cite{Dobaczewski1995}. However, these relations
do not connect spin and spatial degrees of freedom as consistently as Lorentz
invariance and, in addition, there is ambiguity, because many of these very
successful functionals, still in use, are adjusted without taking them into
account.

 As shown in Ref.~\cite{Hofmann1988} for fields of the
form~(\ref{A-ext})
all induced currents and magnetic potentials are parallel to $\boldsymbol{e}%
_{\varphi}$ and axial symmetry is preserved as a self-consistent
symmetry~\cite{RS.80}. One can write out explicitly the Klein-Gordon
equations
for the time- and space-like meson fields as
\begin{align}
\left(  -\Delta+m_{\sigma}^{2}\right)  \sigma &  =-g_{\sigma}\left(  \rho
_{s}^{p}+\rho_{s}^{n}\right)  -g_{2}\sigma^{2}-g_{3}\sigma^{3},\nonumber\\
\left(  -\Delta+m_{\omega}^{2}\right)  \omega_{0}  &  =g_{\omega}\left(
\rho_{v}^{p}+\rho_{v}^{n}\right)  ,\nonumber\\
\left(  -\Delta+m_{\omega}^{2}\right)  \boldsymbol{\omega}  &  =g_{\omega
}\left(  \boldsymbol{j}^{p}+\boldsymbol{j}^{n}\right)  ,\nonumber\\
\left(  -\Delta+m_{\rho}^{2}\right)  \rho_{0}  &  =g_{\rho}\left(  \rho
_{v}^{p}-\rho_{v}^{n}\right)  ,\nonumber\\
\left(  -\Delta+m_{\rho}^{2}\right)  \boldsymbol{\rho}  &  =g_{\rho}\left(
\boldsymbol{j}^{p}-\boldsymbol{j}^{n}\right)  ,\nonumber\\
-\Delta A_{0}  &  =e\rho_{v}^{p},\nonumber\\
-\Delta\boldsymbol{A}  &  =e\boldsymbol{j}^{p}, \label{meseq2}%
\end{align}
where the source scalar and vector densities read
\begin{align}
\rho_{s}^{n,p}  &  =\sum_{i=1}^{N,Z}\psi_{i}^{\dagger}\beta\psi_{i}%
,\nonumber\\
\rho_{v}^{n,p}  &  =\sum_{i=1}^{N,Z}\psi_{i}^{\dagger}\psi_{i}, \label{deneq}%
\end{align}
and the source currents
\begin{equation}
\boldsymbol{j}^{n,p}=\sum_{i=1}^{N,Z}\psi_{i}^{\dagger}\boldsymbol{\alpha}%
\psi_{i}, \label{cureq}%
\end{equation}
where $n$ and $p$ refer to neutrons and protons, respectively. Equations
(\ref{direq}) and (\ref{meseq}) provide a closed set. Their solution has to be
found iteratively, starting from a reasonable estimate of the meson fields,
the Dirac equation (\ref{direq}) is solved yielding the single-particle
spinors. From the spinors, using (\ref{deneq}) and (\ref{cureq}), one obtains
the densities and currents, which act as sources for the solution of the
Klein-Gordon equations (\ref{meseq2}) that provide a new set of meson fields.
Repeating the procedure until convergence results in the self-consistent
solution of this set of equations (see Ref.~\cite{Gambhir1990} for details).
From
this solution one can calculate physical quantities such as the total energy,
radii and deformations.

The actual numerical solution of these coupled set of equations is obtained
using an oscillator expansion in $N=20$ major shells, for which further
technical details can be found in Ref.~\cite{Gambhir1990}. Details pertaining
the
new terms involved in the inclusion of the coupling to an external magnetic
field can be found in the Appendices A and B.

\section{Effects of the magnetic field on the nuclear structure}

The effects that the coupling of protons and neutrons to an external magnetic
field has on the nucleus can be classified as:

\textit{Neutron paramagnetism}: or Pauli-type magnetism, is caused by the
interaction of the magnetic field with the neutron magnetic dipole moment. It
induces a relative shift of levels with neutron spins directed along the
magnetic field. Since the gyromagnetic factor for neutrons is negative ($g_{n}
= -3.8263$), configurations with the spin anti-parallel to the magnetic field
are energetically favored.

\begin{figure}[ptb]
\includegraphics{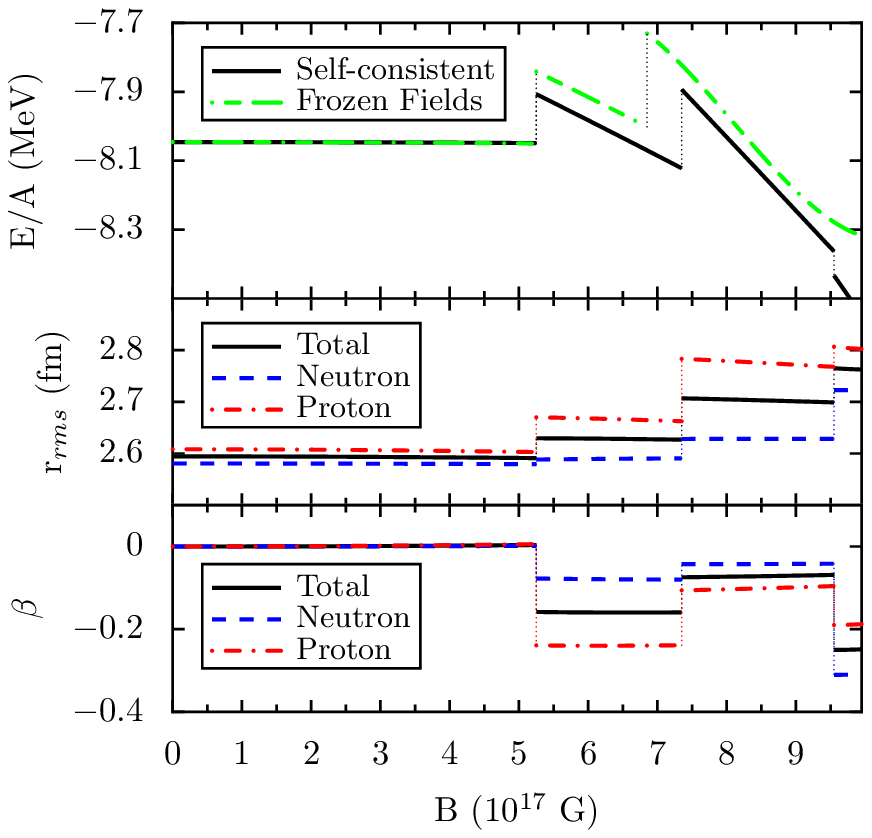} \caption{(Color online) Binding energy per article,
radius and $\beta$ deformation dependence on the magnetic field strength for
$^{16}$O. The binding energy per particle also shows the differences between
using a self-consistent approach and adding the magnetic field in a
frozen-field configuration on top of the bare self-consistent ground-state
(see text for details).}%
\label{f:01}%
\end{figure}

\textit{Proton paramagnetism}: as in the case of neutrons, it comes from the
interaction of the magnetic field with the proton magnetic dipole moment.
However, the gyromagnetic factor for protons is positive ($g_{p} = 5.5856$),
which favors configurations where the proton spin is parallel to the magnetic
field.

\textit{Proton orbital magnetism}: or Landau-type magnetic response, that
couples the orbital motion of protons with the magnetic field. It favors
configurations where the proton angular momentum projection is oriented along
the direction of the external magnetic field.

From the single-particle level point of view, there are two different effects.
The orbital magnetism associated with proton ballistic dynamics removes
Kramer's degeneracy in angular momentum projection $\Omega$ of proton levels,
and brings those aligned with the magnetic field down in energy. On the other
hand, the paramagnetic response (Pauli magnetism) removes the angular momentum
projection degeneracy for both protons and neutrons. It is thus expected that
the magnetic field effect on the single-particle structure is more pronounced
for protons than for neutrons.

\begin{figure}[ptb]
\includegraphics{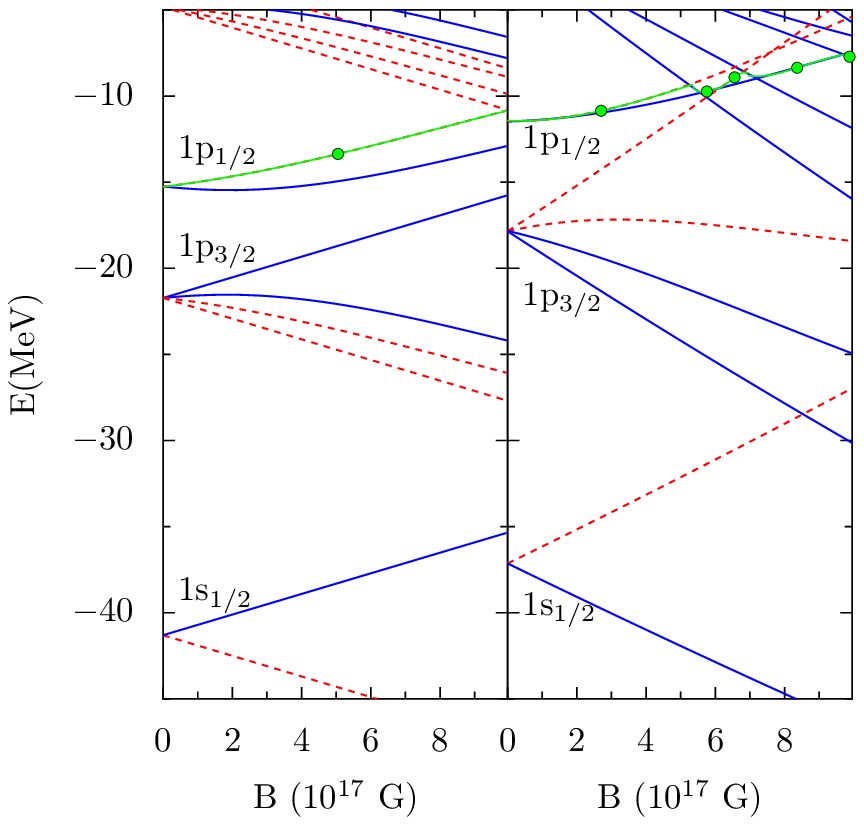} \caption{(Color online) Left: neutrons, right: protons.
Evolution of the single-particle levels in $^{16}$O with increasing
external magnetic field, with frozen nuclear potentials at their values for
vanishing $B$.
 Landau coupling for the protons and
the coupling of the anomalous magnetic moments is included. Blue
lines refer to levels with positive angular momentum projection
$\Omega$, while red lines to levels with negative $\Omega$. Solid
lines indicate positive parity, while dashed lines indicate negative
parity. The green dots mark the last occupied level, while the
magenta circle on the
proton graph highlights the first level crossing at the Fermi energy.}%
\label{f:02}%
\end{figure}

\begin{figure}[ptb]
\includegraphics{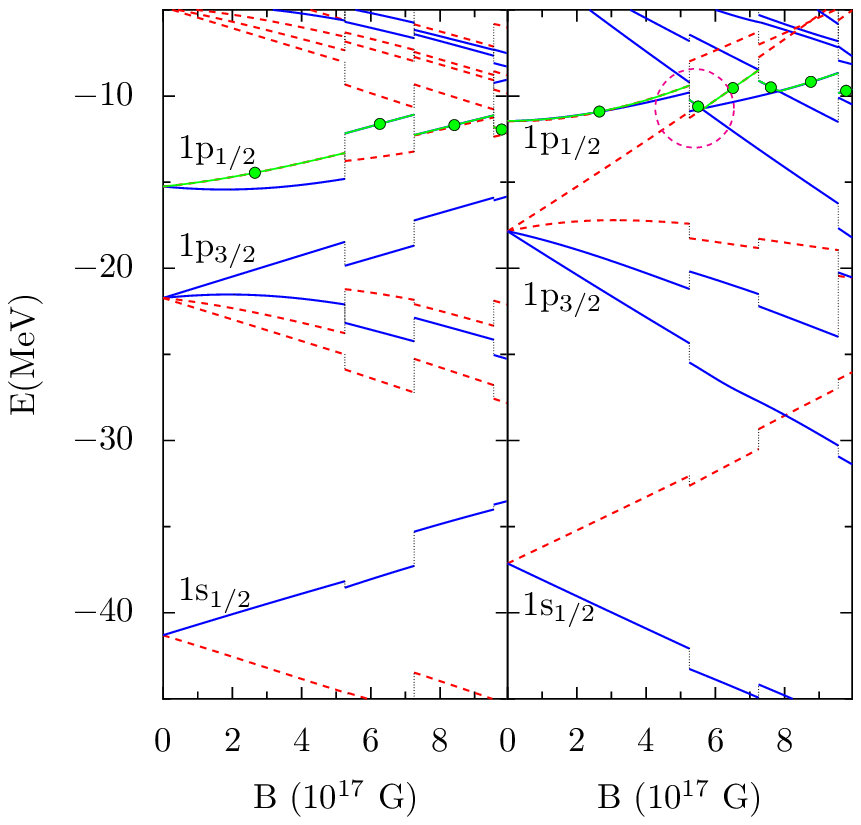}
\caption{(Color online) Same as Fig.
\ref{f:02}, but with a fully self-consistent solution of the equations of
motion (see text for details).}%
\label{f:03}
\end{figure}

As a first step, it is enlightening to study the effects of magnetic
fields on the nuclear structure in a simple and well known nucleus
like $^{16}$O. Fig. \ref{f:01} shows the evolution of the bulk
properties of $^{16}$O with increasing magnetic field. At first the influence
of the external magnetic field is counteracted by the currents generated by
the breaking of time reversal symmetry, including the classical $\propto B^{2}$
contribution coming from the orbital coupling.
The radius of the nucleus
and spherical shape show high resilience to the increase in
the external magnetic field. For field strengths around $5 \cdot
10^{17}$ G, there is an abrupt decrease in binding energy, associated
with increased radius and the sudden appearance of oblate deformation
for the ground state. Such discontinuities are an indication
that the underlying shell structure has changed in a fundamental way.
These jumps in bulk nuclear properties can be traced to the
single-particle behavior, as can be observed in Figs. \ref{f:02} and
\ref{f:03}; they occur when the last occupied level crosses the first
empty level. At that point, for even-even nuclei, a reoccupation
occurs. A particle is removed from a level going upwards with
increasing B-field and brought to a level going downward with
increasing spin. Since these two levels have opposite angular
momentum along the symmetry axis, the nucleus becomes spin-polarized.
Another effect on the structure of nuclei which are superfluid for
vanishing external magnetic field is the gradual disappearance of the
neutron and proton pairing-gaps with increasing external field. The
original shell structure is washed out due to the complicated pattern
of level crossings, and, as the magnetic field increases, new magic
numbers may appear.

In the upper panel of Fig.~\ref{f:01} we show two types of
calculations. The full curve correspond to self-consistent
calculations, where the nuclear potentials change with increasing
external fields due to polarization of the densities and due to
the polarization currents (nuclear magnetism). The dashed curve (frozen
potentials) correspond to a calculation where, in a first step, the
nuclear potentials $S({\bf r})$ and $V_0({\bf
r})=g_{\omega}\omega_0({\bf r})+g_{\rho}\tau_{3}\rho_{0,3}({\bf
r})+eA_{0}({\bf r})$ are calculated without external magnetic field
and subsequently these potentials are kept frozen while the external
magnetic field
is switched on. Therefore, in this case we have no nuclear magnetism.
For small external
fields, i.e. up to the first level crossing at $B\approx 5.2\cdot
10^{17}$ G there is practically no difference. Levels
with $\pm\Omega$ are equally occupied and the corresponding single
particle wave functions are very similar; their
contributions to the currents nearly cancel each other and there is
practically no polarization and no nuclear magnetism. The situation
changes, however, after the first level crossing. Now the nucleus is
spin polarized in the self-consistent solutions (full curve), we
have polarization currents and nuclear magnetism, effects neglected
in the calculations with frozen fields. Thus we observe
differences in the binding energies and also in the location of the
next level crossings.

In order to understand the results of Fig.~\ref{f:01} in more detail
we consider in the Figs.~\ref{f:02} and~\ref{f:03} the effects on the
single-particle structure outlined at the beginning of this section
for frozen fields and for self-consistent fields. Proton and neutron
level degeneracy is broken in reversed directions due to the
different sign in their paramagnetic behavior interacting with the
external magnetic field. This degeneracy breaking is more acute in
the case of protons, where the orbital magnetism plays an important
role. Therefore the first level crossing occurs for the protons at
$B\approx 5 \cdot 10^{17}$ G. For frozen fields (Fig.~\ref{f:02})
this first level crossing for protons has no influence on the other
proton levels nor on the neutron levels. This is no longer the case
for the self-consistent solution in Fig.~\ref{f:03}, where the
changes in the nuclear fields caused by polarization are clearly seen
also in the other proton levels and due to the proton-neutron
interaction also in the neutron levels.

Polarization effects induced by the
$\omega$ and $\rho$ currents due to breaking of time-reversal
symmetry are important, and the frozen fields approximation breaks
down for higher magnetic fields. In the first level crossing one
proton is removed from the $1p_{1/2}$ shell and brought to the
downward sloping orbit of the $1d_{5/2}$ shell, i.e. to a level with
$\Omega=+5/2$. In the second and third panels of Fig.~\ref{f:01}
we see that the reoccupation corresponds to an increase of the proton
radius and transition from a spherical shape to an oblate
deformation.

Light stable nuclei, like $^{16}$O in Fig.~\ref{f:01}, are very stiff
in their response to the external magnetic field. In this particular
case the change in binding energy is less than 100 keV per particle
for field strengths less than $5 \cdot10^{17}$ G. The induced
currents tend to counteract the effects of the magnetic field. The
point at which the first level crossing occurs may be arbitrarily
defined as the minimum field strength for which the nuclear structure
is significantly altered. Thus for $^{16}$O that would be $5
\cdot10^{17}$ G, with a jump in binding energy per particle of around
400 keV. For less stable and/or heavier nuclei, as will be studied in
the next section, it is expected that this minimum field is reduced,
due to, mainly, two effects: i) increase in the level density around
the Fermi energy, and ii) the increase in the proton orbital magnetic
response due to the occupation of single-particle orbitals with
higher angular momenta. In particular, this implies that the validity
of a frozen-field treatment of the coupling to an external magnetic
is highly dependent on the nucleus under consideration. In the next
section it shall be investigated the response of heavier nuclei to
the external magnetic field.

\section{Possible influence on the outer crust composition}

The outer crust of neutron stars, below the neutron drip density, is believed
to be composed by well separated nuclei positioned on a body centered cubic
(bcc) lattice in complete thermodynamical equilibrium. Assuming that matter in
such conditions condenses to a perfect crystal lattice with a single nuclear
species at each site, the energy density is \cite{Haen89p353,Lai1991}
\begin{equation}
\epsilon= n_{N} W_{N}(A,Z) + \epsilon^{\prime}_{e}(n_{e}) + \epsilon
_{L}(Z,n_{e}) \label{enerdens}%
\end{equation}
where $W_{N}$ is the mass-energy of the nuclear species, $\epsilon^{\prime
}_{e}$ is the free energy of the electrons, $\epsilon_{L}$ is the bcc Coulomb
lattice energy, and $n_{N}$ and $n_{e}$ are the number densities of nuclei and
electrons. At a given pressure, minimization of the Gibbs free energy per
nucleon with respect to the nuclear species (A,Z) determines the nuclear
composition of the lattice vertices.

\begin{figure}[ptb]
\includegraphics{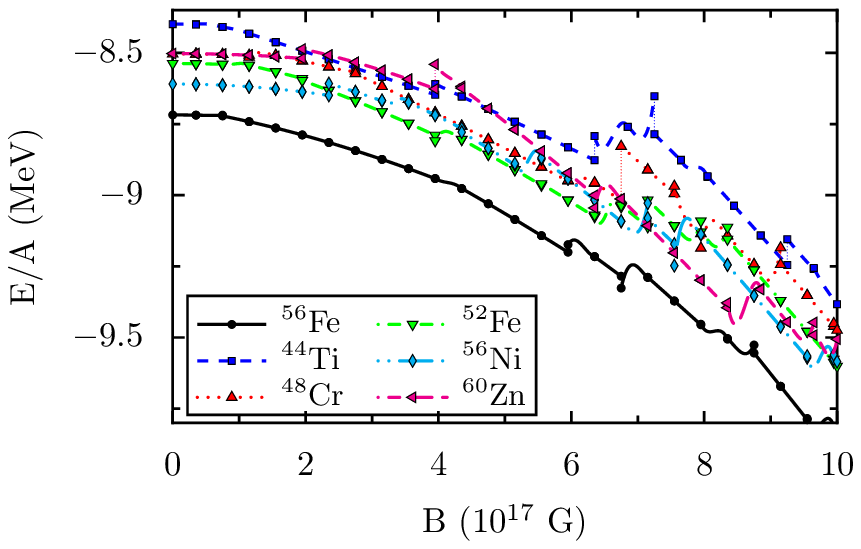} \caption{(Color online) Energy per particle for N=Z
nuclei around $^{56}$Fe.}%
\label{f:04}%
\end{figure}

So far all previous studies (see, for example, [3,27-31]) on the
influence of strong magnetic fields in the composition of the crust have
concentrated on the electron part, and assumed that the nuclear binding energy
$W_{N}$ is not affected. This is certainly true for weaker magnetic fields are
below $10^{15}$ G, since such fields do not alter the nuclear structure.
However, previous studies \cite{Kondratyev2000,Kondratyev2001} where the
influence of the magnetic field on the shell correction energy has been
investigated in a simple model, found that fields with a strength above that
threshold might change the nuclear binding energy, and thus modify the
equilibrium nuclear species on the lattice. In fact, changes of a few keV per
nucleon might significantly alter the composition, making more neutron-rich
nuclei dominate over the most likely to be found nuclei around the $^{56}$Fe
region \cite{Goriely2010}.

One important question is the location of this magnetic field
strength threshold, in particular for nuclei in the vicinity of $^{56}$Fe.
Assuming an electron fraction close to $Y_{e}=0.5$ \cite{Woos02p1015}, it can be
argued that the most probable nuclei are those with the same number of protons
and neutrons. Fig.~\ref{f:04} shows the dependence of the nuclear binding energy
per nucleon on the magnetic field strength for N=Z nuclei for fully
self-consistent solutions of the RME equations in Eqs.~(\ref{direq})
and~(\ref{meseq})
in the region close to $^{56}$Fe. For field strengths of $0.5 \cdot10^{17}$ G
there are changes of a few tenths of keV in the binding energy per nucleon of
some species. At higher magnetic fields, around $2 \cdot10^{17}$ G, the
hierarchy in binding energy of the most bound nuclei is altered and thus it is
expected that the composition of the outer crust is substantially altered.

\begin{figure}[ptb]
\includegraphics{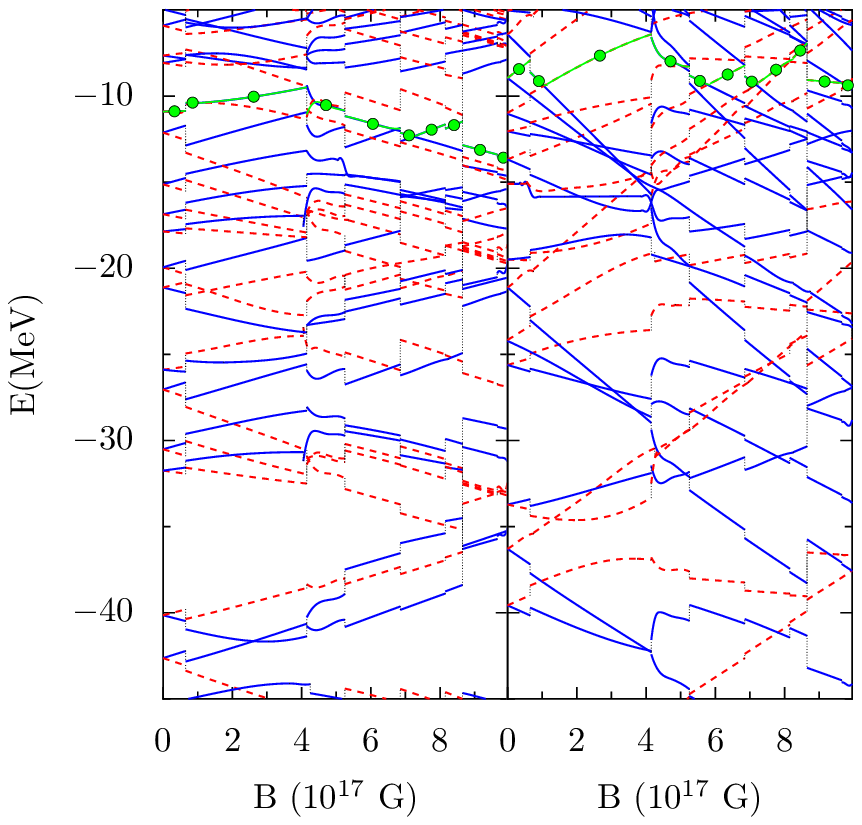} \caption{(Color online) Left: neutrons, right: protons.
Evolution of the single-particle levels in $^{56}$Fe with the magnetic field,
including the proton orbital coupling and the anomalous magnetic moments
coupling. Solid (blue) lines refer to levels with positive $\Omega$, while
dashed (red) lines to levels with negative $\Omega$. Solid lines indicate
positive parity, while dashed lines indicate negative parity. The (green) dots
mark the last occupied level. }%
\label{f:05}%
\end{figure}

As in the example with $^{16}$O, the origin of these discontinuities
in the binding energy per particle can be traced back to the
single-particle structure. Fig.~\ref{f:05} shows the single-particle
level spectra with respect to the external field for $^{56}$Fe. The
diagram is very similar to that of $^{16}$O, only the larger level
density around the Fermi energy and smaller single particle gap
reduce the minimum field for which the magnetic field produces
structural changes. Similar diagrams are found for all of the other
nuclei presented in Fig.~\ref{f:05}, and in those in the vicinity of
$^{56}$Fe. Or course, details pertaining the minimum field that
induces a different single-particle level occupation scheme depend
very much on each particular nucleus.


Concerning the possible changes in the hierarchy of binding energy
per particle with increasing magnetic field, it has been found in this work
that several nuclei (e.g.~$^{57}$Fe and $^{55}$Fe) overbind $^{56}$Fe for
extended ranges of external magnetic field strengths. In that regard, the
frozen field solutions show a different behavior than the self-consistent
ones. The field ranges for which this hierarchy changes occur are
different and the magnitude is typically off by a couple hundred keV
as compared with the fully self-consistent solutions. This ordering
according to binding energy is one of the factors that influences the
final composition of the outer crust in neutron stars, and quantitative
predictions should be done using the full self-consistent formulation.

\begin{figure}[ptb]
\includegraphics{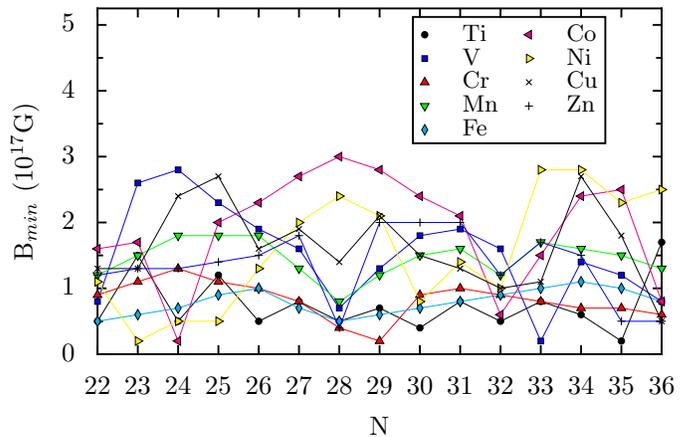} \caption{(Color online) Minimum magnetic field for which
the first level crossing at the Fermi energy occurs for isotopic chains around
$^{56}$Fe.}%
\label{f:07}%
\end{figure}

It is also interesting to study the minimum field (defined as the
field at which the first level crossing occurs) for isotopic chains
in the iron region, since it provides an indication of the possible
effects on the neutron star composition. Fig.~\ref{f:07} shows the
magnetic field at which the first level crossing occurs, for
different isotopic chains close to iron. It provides an intuitive
idea of how intense the magnetic fields have to be on average in
order to affect the nuclear structure significantly. For isotopes
close to $^{56}$Fe this value is approximately between
$0.5\cdot10^{17}$ G and $3\cdot10^{17}$ G. For heavier as well as
neutron-rich nuclei, which exist at higher densities in the crust, a
sharp decrease of this minimum field is expected as mentioned
previously. Thus, for fields around $10^{17}$ G it is reasonable
to expect changes in the crust composition. However, and
because of the strong dependence of this minimum field on the nuclear
species, it is not possible to predict the effect on the composition
without actually performing the minimization of Eq.~(\ref{enerdens}).
This calculation is, however, computationally very demanding and
would also require refinements in the model like the inclusion of
pairing (even though it is reasonable to expect that magnetic fields
damp and eventually cancel it) or the proper inclusion of one-pion
exchange terms in the effective Lagrangian. Therefore, it is well
outside of the scope of the present work.

\section{Conclusions}

The influence of strong magnetic fields on nuclear structure has been
studied using a fully self-consistent covariant density functional.
It has been found that a field strength of at least $10^{17}$ G is
needed to appreciably modify the nuclear ground state. For
sufficiently high fields these effects cannot be studied using a
frozen-field approach since there are level rearrangements, causing
spin-polarization and induced currents and thus a self-consistent
model is required. It is the advantage of covariant theories that the
these currents can be taken into account without any additional
parameters. The minimum magnetic field that changes the nature of the
nuclear ground-state is very much dependent on the nucleus, and its
effects on the nuclear binding energy per nucleon range from a few
tenths of keV for $B\approx0.5\cdot10^{17}$G to a few hundreds of
keV for the maximum field theoretically possible $B \approx10^{18}$
G.

No neutron star has yet been observed with such intense magnetic
fields $B>10^{16}$ G, even though theoretical models hint that such
objects exist \cite{Duncan1997,Thompson2001}. In such a case, the composition of
its outer crust might be radically different from that of normal
neutron stars. The relevance of this change in composition depends on
the abundance of normal neutron stars compared with that of
magnetars. Until now, observations suggest that magnetars are not so
common in the universe and thus it is unlikely that the magnetic
field effects on nuclear structure play an important role in global
astrophysical observables like element abundances. However, changes
in the composition of magnetars might be relevant in the study of
different phenomena in these particular kind of neutron stars, for
example elastic properties of the crust \cite{Pethick1998}, pulsar
glitches \cite{Mochizuki1995}, or cooling \cite{DeBlasio1995}. With the
inclusion
of a proper pairing interaction, the covariant DFT model presented in
this work can be used to perform a quantitative exploration of all these
questions. A systematic study of changes in the composition of the
outer crust will be presented in a upcoming publication.

\vskip 0.5cm

\noindent\textbf{Acknowledgments} D.P.A. is grateful M. Urban, M. Bender, and to
S. Goriely  for very fruitful discussions and the hospitality of the IAA group
in Brussels, where part of this work was done. E.K. is grateful to N. Chamel for
introducing him in the topic of this work. This paper has been supported in part
by the ANR NEXEN and the DFG cluster of excellence \textquotedblleft Origin
and Structure of the Universe\textquotedblright\ (www.universe-cluster.de).

\section{Appendix A: Oscillator matrix elements for the proton orbital
coupling}

\subsection{Dirac equation}

The Dirac equation (\ref{direq}) together with the meson field equations
(\ref{meseq}) are easily solved in a harmonic oscillator basis. The procedure
is described in great detail in Ref. \cite{Gambhir1990}. The only new term
coming
in the solution of the Dirac equation is $\boldsymbol{\alpha} \cdot
\boldsymbol{V}$ where $\boldsymbol{V}$ is given in Eq.~(\ref{e:14}). Using the
set of $\alpha$ matrices in the spherical tensor basis, $( \alpha_{+},
\alpha_{-}, \alpha_{3} )$ with
\begin{equation}
\alpha_{+} = \left(
\begin{array}
[c]{cc}%
0 & \sigma_{+}\\
\sigma_{+} & 0
\end{array}
\right)  , \quad\alpha_{-} = \left(
\begin{array}
[c]{cc}%
0 & \sigma_{-}\\
\sigma_{-} & 0
\end{array}
\right)  ,
\end{equation}
allows to write
\begin{equation}
\boldsymbol{\alpha} \cdot\boldsymbol{V} = \alpha_{+} V^{-} + \alpha_{-}V^{+} +
\alpha_{3} V^{3},
\end{equation}
with
\begin{equation}
\boldsymbol{A}^{(e)} = i \left(  -\frac{rB}{2} e^{-i\varphi},\frac{rB}{2}
e^{i\varphi}, 0 \right)  ,
\end{equation}
and the internal self-consistent magnetic potential $\boldsymbol{V}$ given by
the solution of the Klein-Gordon equations. Using an oscillator expansion for
the spinors
\begin{equation}
\psi_{i} = \left(
\begin{array}
[c]{c}%
\sum_{n} f^{i}_{n} |n\rangle\\
i \sum_{n^{\prime}} g^{i}_{n^{\prime}} |n^{\prime}\rangle
\end{array}
\right)  ,
\end{equation}
where $|n\rangle$ are the axially symmetric harmonic oscillator wave
functions, determined by the quantum numbers $n \equiv(n_{z},n_{r},m_{l}%
,m_{s})$.
\begin{equation}
\vert n\rangle\equiv\vert n_{z} n_{r} m_{l} m_{s} \rangle= \phi_{n_{z}}(z)
\phi^{|m_{l}|}_{n_{r}}(r) \frac{e^{im_{l}\varphi}}{\sqrt{2\pi}}\chi_{m_{s}}(s)
\end{equation}
with $\Omega= m_{l} + m_{s}$. The term $\boldsymbol{\alpha} \cdot
\boldsymbol{V}$ in the Dirac equation is then in matrix form
\begin{equation}
\left(
\begin{array}
[c]{cc}%
0 & \mathcal{B}_{nn^{\prime}}\\
\mathcal{B}_{n^{\prime}n} & 0
\end{array}
\right)  \left(
\begin{array}
[c]{c}%
f_{n}\\
ig_{n^{\prime}}%
\end{array}
\right)  ,
\end{equation}
with
\begin{equation}
\mathcal{B}_{nn^{\prime}} = -\langle n\vert V^{-} \sigma_{+}\vert n^{\prime
}\rangle- \langle n\vert V^{+} \sigma_{-}\vert n^{\prime}\rangle.
\end{equation}
The oscillator matrix elements can be written as
\begin{align}
\langle n\vert V^{-} \sigma_{+}\vert n^{\prime}\rangle= + \delta_{m^{\prime
}_{l},m^{}_{l}+1} \langle n^{}_{z} n^{}_{r} m^{}_{l} \vert V^{-}\vert
n^{\prime}_{z} n^{\prime}_{r} m^{\prime}_{l} \rangle\\
\langle n\vert V^{+} \sigma_{-}\vert n^{\prime}\rangle= - \delta_{m^{\prime
}_{l},m^{}_{l}-1} \langle n^{}_{z} n^{}_{r} m^{}_{l} \vert V^{+}\vert
n^{\prime}_{z} n^{\prime}_{r} m^{\prime}_{l} \rangle.
\end{align}

\subsection{Currents in coordinate space}

To solve the Klein-Gordon equations, the source terms have to be transformed
from oscillator to coordinate space. Expressions for the scalar and vector
densities are given in Ref.~\cite{Gambhir1990}. The currents are defined as
\begin{equation}
\boldsymbol{j}(\boldsymbol{r}) = \sum_{i} \psi_{i}^{\dagger}(\boldsymbol{r})
\boldsymbol{\alpha} \psi_{i}(\boldsymbol{r})
\end{equation}
Because of axial symmetry the third component vanishes, and we have to
consider only two of the components:
\begin{equation}
j_{+}(\boldsymbol{r}) = \sum_{i} \psi_{i}^{\dagger} \alpha_{+} \psi_{i} \qquad
j_{-}(\boldsymbol{r}) = \sum_{i} \psi_{i}^{\dagger} \alpha_{-} \psi_{i}.
\end{equation}
Since $\alpha_{+} = (\alpha_{-})^{\dagger}$ we have $j_{-} = j_{+}^{*} $, so
only one of them needs to be calculated explicitly. With the Dirac spinors in
Eq.~(\ref{def:wavef}) we find
\begin{align}
j_{+}(\boldsymbol{r})  &  = + i e^{-i\varphi} j(r,z)\\
j_{-}(\boldsymbol{r})  &  = - i e^{+i\varphi} j(r,z)
\end{align}
with
\begin{align}
j(r,z)  &  = \sum_{i} \sum_{nn^{\prime}} f^{i}_{n} g^{i}_{n^{\prime}} \Phi
_{n}(r,z) \Phi_{n^{\prime}}(r,z) \delta_{ m^{\prime}_{l},m^{}_{l}+1 }\\
&  - \sum_{i} \sum_{nn^{\prime}} f^{i}_{n}g^{i}_{n^{\prime}} \Phi_{n}(r,z)
\Phi_{n^{\prime}}(r,z) \delta_{ m^{\prime}_{l},m_{l}-1 }\nonumber
\end{align}
where the $\Phi_{n}(r,z)$ are the oscillator wave functions without spin
dependence, i.e. $\Phi_{n}(r,z) = \phi_{n_{z}}(z)\phi_{n_{r}}^{|m_{l}|}(r)$.

\subsection{Klein-Gordon equation oscillator matrix elements for the vector
terms}

In the spherical tensor basis, the K-G equations read
\begin{equation}
( - \Delta+ m^{2} ) w^{i} = g j^{i} \qquad i = +, -, 3
\end{equation}
for the massive mesons. The functional form of the currents, $j_{\pm}%
(r,\theta,z) = \pm i \, j(r,z) e^{\mp i\theta}$, suggest the following ansatz
for the potentials: $w_{\pm}(r,\theta,z) = \pm i \, w(r,z) e^{\mp i\theta}$.
Inserting both in the K-G eqs, and eliminating the angular dependence in
$\theta$
\begin{equation}
\left(  - \partial^{2}_{r} - \frac{1}{r}\partial_{r} - \partial^{2}_{z} +
\frac{1}{r^{2}} + m^{2} \right)  w(r,z) = j(r,z)
\end{equation}
The oscillator matrix elements for the Laplacian can be found in
\cite{Gambhir1990}. It is only a matter of including the oscillator matrix
elements
for the $1 / r^{2}$ term, which can be accomplished trivially in coordinate
space.

\section{Appendix B: Oscillator matrix elements for the coupling with
intrinsic magnetic moments}

\label{appendixB}

The coupling of the magnetic field with the anomalous magnetic moments of
protons and neutrons introduces a new term in the single-particle Dirac
equation (\ref{direq}) with
\begin{equation}
\chi_{\tau_{3}}^{(e)}=\kappa_{\tau_{3}}\mu_{N}\frac{1}{2}\sigma_{\mu\nu
}F^{(e)\mu\nu}%
\end{equation}
For a constant magnetic field $\boldsymbol{B}$ of the form (\ref{A-ext}) it is
easy to show that
\begin{equation}
\frac{1}{2}\sigma_{\mu\nu}F^{(e)\mu\nu}=-\left(
\begin{array}
[c]{cc}%
\boldsymbol{\sigma} & 0\\
0 & \boldsymbol{\sigma}%
\end{array}
\right)  \cdot\boldsymbol{B}=-\boldsymbol{\Sigma}\cdot\boldsymbol{B},
\end{equation}
and for $\boldsymbol{B}=(0.0,B)$ we have
\begin{equation}
\chi_{\tau_{3}}^{(e)}=-\kappa_{\tau_{3}}\mu_{N}\Sigma_{3}B,
\label{sigmaB}
\end{equation}
Finally, the oscillator matrix elements can be written as
\begin{equation}
\langle n|i\chi_{\tau_{3}}^{(e)}|n^{\prime}\rangle=\pm\delta_{n_{z}%
n_{z}^{\prime}}\delta_{n_{r}n_{r}^{\prime}}\delta_{m_{l}m_{l}^{\prime}}%
\delta_{m_{s}m_{s}^{\prime}}\kappa_{\tau_{3}}\mu_{N}B\text{ \ \ }%
\end{equation}
\ for \ $m_{s}=\pm\frac{1}{2}$.

\end{document}